\documentclass[british]{article}
\usepackage[T1]{fontenc}
\usepackage[latin9]{inputenc}
\usepackage{geometry}
\usepackage{prettyref}
\usepackage{float}
\usepackage{graphicx}

\makeatletter

\providecommand{\tabularnewline}{\\}

\newenvironment{lyxlist}[1]
{\begin{list}{}
{\settowidth{\labelwidth}{#1}
 \setlength{\leftmargin}{\labelwidth}
 \addtolength{\leftmargin}{\labelsep}
 }}
{\end{list}}

\usepackage{prettyref}
\newrefformat{tab}{Table\,\ref{#1}}
\newrefformat{fig}{Figure\,\ref{#1}}

\makeatother

\usepackage{babel}
\begin{document}

\title{Image compression overview}

\date{M. Prantl%
\thanks{info@perry.cz%
}}

\maketitle
Compression plays a significant role in a data storage and a transmission.
If we speak about a generall data compression, it has to be a lossless
one. It means, we are able to recover the original data 1:1 from the
compressed file. Multimedia data (images, video, sound...), are a
special case. In this area, we can use something called a lossy compression.
Our main goal is not to recover data 1:1, but only keep them visually
similar. This article is about an image compression, so we will be
interested only in image compression. For a human eye, it is not a
huge difference, if we recover RGB color with values {[}150,140,138{]}
instead of original {[}151,140,137{]}. The magnitude of a difference
determines the loss rate of the compression. The bigger difference
usually means a smaller file, but also worse image quality and noticable
differences from the original image. 

We want to cover compression techniques mainly from the last decade.
Many of them are variations of existing ones, only some of them uses
new principes.

\section{Introduction }

Someone may ask the important question ``Why compress data at all''.
Well, lets dive into some facts. Single image at 1920x1080 resolution
(with RGB colors) gives us the uncompressed (raw) size of 6075KB for
a single image. Many of the digital cameras takes even larger pictures.
To keep them uncompressed would be a waste of space and also energy
(a bigger storage space consumes more electricity and produce more
heat, that needs to be cooled down with another energy etc.).

The lossless compression is suitable for images, that will be later
edited or need to keep fine details. Sometimes, we can accept a small
change in a compressed data. Those methods are called near-lossless
and are often used in medical images. The largest group is covered
by a lossy compression. 

The quality of the compression can be expressed by the compression
ratio. For lossless methods, we can get the average of 3-4 times smaller
files than the original ones. With lossy methods, we can obtain ratios
up to 50:1 while maintain good perceptual quality of a reconstructed
data. Rating factor of compression is not only in its efficiency in
term of compression ratio. Very significant is also a compression
time complexity. Many compression algorithms are asymmetric, which
means that compression takes more time than decompression. That is
usually not problem for images, since we compress file only once,
but decompress it repeatedly. This could cause some problems in real
time applications.

There are many different approaches for compression of different data
types. Its not possible to cover all of them. In this paper we are
going to focus on compressions of common images. We are not considerating
special methods based on the prior knowledge of data types, such as
medical images, satellite images, GIS. We decided to take wider range
of approaches and each of them support with small example of existing
algorithms. Our main interest is in the last ten years of research.
Both, lossless and lossy methods are discussed.

\section{Lossless methods}

Lossless methods are very general. In theory, we can use any lossless
compression algorithm and apply it to images. In practice, that is
not usually the best idea. There are several approaches, that are
designed to be more efficient with images. Compressions can be divided
into several categories, based on algorithm main idea. There are three
main directions \textendash{} dictionary, prediction and wavelet based
methods.

\subsection{Dictionary based }

Those algorithm are based on a dictionary methods. Most of them are
variations of LZ-family algorithms (LZ77, LZ78, LZSS, LZW...) \cite{Sal00}.
Main idea comes from LZ77 \cite{Ziv77} algorithm. Superior version
further improved compression ratio.

LZ77 is working with two parts of data at the same time \textendash{}
actual window and sliding window. We are searching for match from
actual window in sliding window. If match is found, reference to the
sliding window is stored. This first version is not optimal, because
it has fixed structure of coded word {[}position, length, next character{]}. 
\begin{lyxlist}{00.00.0000}
\item [{Compression}] example 
\item [{Abracadabra}] = {[}0,0,a{]}{[}0,0,b{]}{[}0,0,r{]}{[}3,1,c{]}{[}2,1,a{]}{[}7,3,a{]} 
\end{lyxlist}
There are two typical, well known, formats, where we can find that
type of compression. Those are PNG \cite{Cro95} and GIF. While GIF
is using LZW (patented), PNG uses open-source variant of original
version of LZ77 algorithm. 

Recently, LZW method in a combination with BCH codes (error correcting
codes) for removal of repeating parts after the compression was proposed
by \cite{Ala12}.

\subsection{Prediction based }

Prediction methods use prediction of a next value based on a special
predictor. There can be different versions of them. Those predictors
are designed to be efficient with image data. Algorithm has two major
steps: 
\begin{enumerate}
\item Prediction of a next value from previous ones. The predicted values
are very close to the original ones 
\item Calculate an error of a prediction as a difference of the original
and the predicted value. Only resulting error is coded, mostly by
entropy coding. 
\end{enumerate}
There are many types of predictors.

\subsubsection{Gradient Adaptive Predictor (GAP) }

Firstly used in CALIC \cite{Wu96}. Its adaptive and non-linear. Due
to its adaptivity its better than classic linear predictors. This
improvement can be best seen along strong edges. Basic equation is
simple and can recognize six types of edge - sharp, classic and weak
in vertical or horizontal direction. Full equation to compute predicted
value can be found in \cite{Wu96}.

\subsubsection{Adaptive Linear Prediction and Classification (ALPC)}

Another adaptive and non-linear predictor was proposed by\emph{ \cite{Mot00}}.
It uses weighted pixel values. Weights can be updated during compression
process to improve prediction accuracy. This predictor is not based
on a single equation, but it uses a simple algorithm to generate weights
for the neighbouring pixels. The pseudo-code of the algorithm is clearly
explained in \cite{Mot00}.

\subsubsection{Median Edge Detection predictor (MED) }

Successor of GAP. Firstly used as a part of LOCO-I algorithm \cite{Wei00}
known from JPEG-LS. The prediction scheme is presented in an equation
\ref{eq:MED}.

\[
\left[\begin{array}{cc}
C & B\\
A & x
\end{array}\right]
\]

\begin{equation}
x_{i+1}=\left\{ \begin{array}{c}
min(A,B)\;\; if\: c\geq max(A,B)\\
max(A,B)\;\; if\: c\leq min(A,B)\\
A+B-C\;\; otherwise
\end{array}\right.\label{eq:MED}
\end{equation}

This predictor was further improved. Predicted errors are decorrelated
in their bit-planes using XOR operator and further reduced with logical
minimization using Quine-McCluskey algorithm. This approach is described
in \cite{Raw12}.

\subsubsection{Neural Network predictor}

Fixed formula based on previous values in not used as a predictor.
Feed forward neural network is used instead. Inputs are based on previous
pixels values and the output is a predicted value. In the learning
phase, the output is set to the correct value. The neural network
is trained using back-propagation algorithm. This method was proposed
by \cite{Ale08}. Similar method with neural predictor was proposed
by \cite{Put11}. This scheme is using larger area than \cite{Ale08},
from which predicted value is calculated.

\subsubsection{Gradient Edge Detection predictor (GED)}

Combination of MED and GAP predictor was proposed by \cite{Avr10}.
Their prediction template uses five neighbouring pixels, as shown
on image X.

Prediction scheme is simple and very similar to MED as we can see
from equation \ref{eq:GED}.

\[
\left[\begin{array}{ccc}
. & . & E\\
. & C & B\\
D & A & x
\end{array}\right]
\]

\[
g_{v}=|C-A|+|E-B|
\]

\[
g_{h}=|D-A|+|C-B|
\]

\begin{equation}
x_{i+1}=\left\{ \begin{array}{c}
A\;\; if\:(g_{v}-g_{h})>T\\
B\;\; if\:(g_{v}-g_{h})<-T\\
\frac{3(A+B)}{8}+\frac{C+D+E}{12}
\end{array}\right.\label{eq:GED}
\end{equation}

\subsection{Wavelet based }

First a discrete wavelet transform (DWT) is applied. This results
in a wavelet coefficients. There is no compression at that moment,
because the count of coefficients is the same as of image pixels.
However, values are more compressible, because they are concentrated
in just few different coefficients. Those coefficients are further
coded using ordering scheme and entropy coding.

Two main ordering algorithms are commonly used \textendash{} Embedded
Zerotrees of Wavelet (EZW) \cite{Sha93} and Set partitioning in hierarchical
trees (SPIHT) \cite{Sai96}. EZW and SPIHT are both progressive algorithm
and uses quad-tree scheme with a predefined scan order of coefficients,
where the most important coefficients are encoded as first. SPIHT
can be seen as an improved version of EZW.

Main disadvantage of DWT is its speed. Compression and decompression
takes usually more time than prediction and dictionary based methods,
but compression ratio results are often comparable. Well known format
based on DWT is JPEG2000 \cite{Rab02}. In \cite{Bra09} better compression
of DWT coefficients via improved SPIHT is proposed. Latest research
proposed by \cite{Li13} further improved compression ratio. There
is also a binary version of this transform. Its faster than full DWT
and used primary for binary images. \cite{Pan07} extend binary transform
for grayscaled images.

\subsection{Other }

Apart from a classic and commonly used algorithms, there can be found
other solutions. Many compressions can be improved by histogram update
in preprocessing step. This improvement was discussed in article \cite{Fer02}.
Method using binary trees and a prediction is described in \cite{Lar12}.
Compression based on a bit-planes and context modelling is discussed
in \cite{Kik09}. 

In 2011 \cite{Mey11} proposed method based on contour compression.
They are traversing along contours with the same pixel brightness
value and store resulting contours using the run length encoding (RLE)
algorithm and an entropy coding.

\section{Near-Lossless methods}

Special category between lossless and lossy methods are near-lossless
ones. They allow a small loss of quality, which is almost non-observable.
Their PSNR (see section 5 for details) is usually very high (50+).
For example, if two neighbouring pixels have values 159, 160, decompressed
values can be 160, 160. Algorithms often proceed from lossless versions,
with some modifications. 

Very suitable are prediction-based methods, where prediction can include
some kind of quality loss. The typical example is MED predictor and
LOCO-I algorithm \cite{Wei00}. At the end, every predictor mentioned
in a previous chapter can be updated to support near-lossless compression. 

A different method for near-lossless mode only based on the prediction
was proposed by \emph{\cite{Nas07}}. Their algorithm uses prediction
and graph of predicted errors (referred as trellis). For each row
of an image, a graph is constructed. Its edges are labelled with the
prediction error. The graph is later processed to found edges that
has the same or similar error value. For those values, run length
encoding (RLE) is used.

Another approach is described in \cite{Mam12}. First, pixels are
divided by 2P (P is number of bits of loss) and rounded. Pixels are
rearranged using different scan orders (Zig-Zag, Snake, Hilbert...).
Only differences between consecutive values are stored and compressed
with the Huffman coding. Best scan order is selected with brute-force,
simple by trying all combinations and choose the best one. 

Attempt to used dictionary based method (LZW) was proposed by\emph{
\cite{Bre03}}. LZW has been modified. Entries in dictionary are searched
with error threshold. There can be more than one result. Selected
entry is the one with minimal error. In decompression, different data
are obtained, but their error is lower than threshold value. Article
also contains comparison with quadtree based near-lossless method.
Quadtree recursive division stop condition has been changed. If block
error is within selected boundaries, no further division occurs and
block is represented by its mean value.

\section{Lossy methods}

\begin{figure}[H]
\begin{centering}
\includegraphics[scale=0.5]{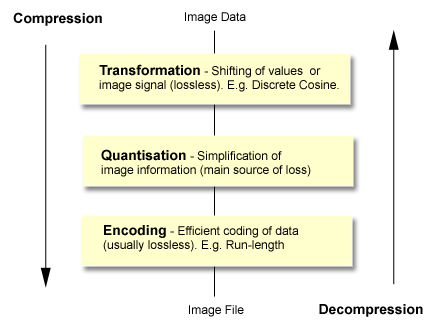}
\par\end{centering}

\caption{Lossy compression / decompression pipeline\label{fig:Comp_pipeline}}

\end{figure}

The lossy methods are of great importance in an image compression.
Human eyes can be tricked and loss of details does not much affect
the overall perception. If we are talking about lossy compressions,
there is always a rate of loss. We can compress the whole image by
only using one dominant color, but what the result will be? Important
is to find a good trade off between the quality and overall image
size. Most of the methods are transform based. Image is transformed
to discrete spectrum and coefficients are quantized and truncated.
Resulted data are losslessly compressed. Pipelined process can be
seen on \prettyref{fig:Comp_pipeline}.

The typical representatives of these methods are discrete cosine transform
(DCT), generalized Karhunen-Loewe transform (KLT) and wavelet based
transform (DWT). Those are basically state-of-the art approaches.

\subsection{Transform based methods }

\subsubsection{DCT }

The most well-known methods are based on the discrete cosine transform
(DCT). It was first presented in \cite{Wal92}. Basic version of the
algorithm is very fast and simple. Image is divided to blocks of size
MxM. Each block is transformed with DCT. Resulting coefficients are
quantized with respect to the loss of quality. Quantized coefficients
are finally compressed with the Huffman or Arithmetic coding. For
its simplicity, a lot of improvements were proposed. \cite{Raj07}
proposed different scheme based on 3D DCT. MxM blocks of image are
grouped to sections of size M, which gives as \quotedblbase{}cube\textquotedblleft{}.
Values are than processed with 3D DCT. The rest of the algorithm is
very similar to the classic DCT. Some research has also been done
in a calculation of DCT matrix. Its approximation for faster compression
and decompression was proposed by \cite{Sen10}. Another approach
from \cite{Cin11} modifies classic matrix and replace its values
with only -1, 0, 1. That leads to faster calculation, because no division
or multiplication is needed. However, result image has a lower visual
quality.

\subsubsection{KLT (PCA) }

Karhunen-Lowe transform is based on a principal component analysis
(PCA). In its core, its same as DCT. The main difference is in a transform
matrix. KLT matrix is obtained from eigen vectors of covariance matrix.
Eigen vectors are reordered with respect to their eigen values. Ordered
vectors are used as rows (or columns) of a transform matrix. More
related information can be found in \cite{Rao00}. DCT can be considered
as averaged KLT. KLT is, unlike DCT, optimal transform. That means,
we can achieve better quality with smaller size. Main disadvantages
comes from transform matrix. This matrix need to be stored along with
a compressed image. Second, its computation can be slow. One solution
to this problem is proposed in \cite{Zha05}.

\subsubsection{DWT }

The main core is almost identical with the one used for lossless compression
(as described before). Transformed values are quantized before final
step (using EZW or SPIHT \cite{Sai96}). Apart integer transformation,
the floating point version can be used for lossy compression. Well
known format based on DWT is JPEG2000 \cite{Rab02}. Its result for
lower bitrates were improved by \cite{Sch09}. Their method use image
division similar to the one used by quadtree based methods. Division
is done based on an error of the block reconstruction. For reconstruction,
colors are interpolated with anisotropic diffusion.

\subsection{Other methods}

\subsubsection{Neural Networks }

\begin{figure}[H]
\begin{centering}
\includegraphics[scale=0.5]{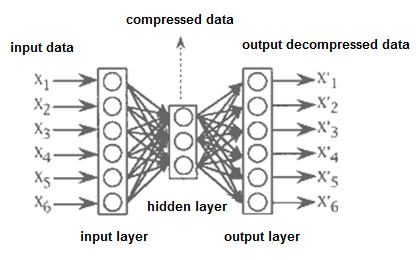}
\par\end{centering}

\caption{Scheme of feed-forward neural network for lossy compression\label{fig:NN}}

\end{figure}

Image can be compressed with an usage of artificial neural networks.
Basic idea is to take a feed-forward neural network with the same
number of inputs and outputs. The hidden layer of a network represents
the compressed data. Simple network scheme can be seen on \prettyref{fig:NN}.
Apart from storing hidden layer, neural network weights must be stored
as well. 

The compression process consist of two main steps. First, the neural
network must be trained to adjust weights, that were randomly selected.
With those adjusted weights, the result hidden layer is generated.
Input image is usually divided into N dimensional vectors. Those vectors
are inputs of an neural network. In the compression step, each of
N vector is replaced with M values from the hidden layer. One of the
main disadvantages is the compression time. Adjusting weights is a
slow process. An improvement of a compression speed was proposed in
\cite{Raj11}. Initialized weights are not randomly selected, but
they are chose as a result from genetic algorithm. Later, weights
are improved with traditional back-propagation. Another approach is
to improve back-propagation \cite{Ira09} or use a different training
method. One possibility is theLevenberg-Marquardt method as proposed
in \cite{Kar11}. Use of this method can be significant speed up for
the networks with only one hidden layer (\cite{Wil99}, \cite{Wil10}).
Improvement in compression ratio and quality can be achieved witch
histogram equalization \cite{Dur05}.

\subsubsection{Contours }

The main idea is to compress only important parts of an image. The
reconstructed image is than very similar to a vector graphics representation
of bitmap image. Level of details in image depends on the compression
ratio. Those method achieves a very good results for a cartoon based
images, where contours are the most important parts. Areas are usually
filled with not much detail. Such method was described by \cite{Mai09}.
Same author in \cite{Mai11} further improved method and test it also
on natural images. They used the Marr-Hidreth detector to obtain image
edges (but any other edge detector can be used as well). Obtained
edges are filtered to preserve only those, that are important. Result
is binary image, losslessly encoded with JBIG algorithm. For color
compression is used quantization scheme. Colors from both sides of
contours are obtained and quantized. Result is again losslessly saved
with PAQ algorithm. Its computational cost is balanced with compression
ratio. For decompression, quantized colors are interpolated with homogeneous
diffusion. For targeting cartoon images, quality is very good (no
artefacts like in transform based methods), but for natural ones,
the quality is worse. The method for natural images was proposed in
\cite{Sca11}. Edges are not obtained with an edge detector, but with
a method similar to marching squares. Its advantage is, that we can
obtain enclosed areas. For every brightness value in the image, set
of contours is obtained. According to the threshold level, some of
them are omitted and rest is stored with its brightness value. Level
of detail can be also controlled. One possibility is to store only
those contours that are longer than a certain threshold. Second possibility
is to simplification using e.g. the Ramer\textendash{}Douglas\textendash{}Peucker
algorithm \cite{Dou73}. A contour can be later stored by its significant
points or using chain code. Chain codes can be more compact.

\subsubsection{SVD}

With an image representation as a matrix in a mathematical sort of
view, work methods based on SVD (Singular Value Decomposition), first
introduced in \cite{And76}. Image is decomposed to three matrices
(equations \ref{eq:SVD1} and \ref{eq:SVD2}). 

\begin{equation}
A=U\Sigma V^{T}\label{eq:SVD1}
\end{equation}

\begin{equation}
A=\left[u_{1}.....u_{n}\right]\left[\begin{array}{ccccc}
\sigma_{1}\\
 & .\\
 &  & .\\
 &  &  & .\\
 &  &  &  & \sigma_{n}
\end{array}\right]\left[\begin{array}{c}
v_{1}\\
.\\
.\\
.\\
v_{n}
\end{array}\right]\label{eq:SVD2}
\end{equation}

\begin{figure}[H]
\begin{centering}
\includegraphics[scale=0.5]{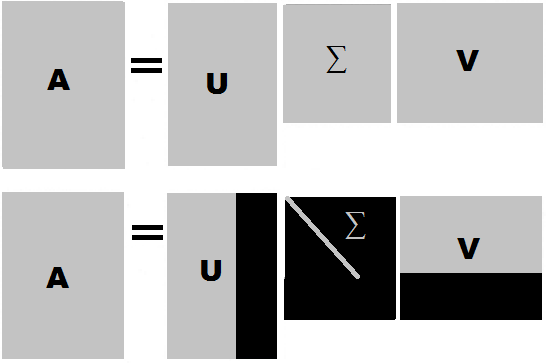}
\par\end{centering}

\caption{Process of SVD compression. First line are full SVD matrices. Second
line represent compression form after removing unnecessary values
in black areas.\label{fig:SVD}}
\end{figure}

There is no compression at this point, since we enlarged original
8bit values to floating point numbers. Matrix \ensuremath{\sum} is
diagonal, U and V members are vectors ($u_{n}$ with size M and $v_{n}$
N for MxN image). Compression can be achieved by removing columns
and rows from matrices, as illustrated on \prettyref{fig:SVD}. 

Main disadvantage is a speed of the compression. Quality is also lower
with comparison to DCT. SVD method was improved for example in \cite{Ran07},
where image is preprocessed before SVD step. In this article can also
be found comparison of SVD with KLT transform. SVD variation for color
images can be found in \cite{Bet05}.

\subsubsection{Other }

There are other methods that were researched. One of such approaches
is based on a spline interpolation of image \cite{Mut08}. Only a
few pixels of image are taken and those are interpreted with a cubic
spline. Control points of a spline are selected to minimize overall
error of approximation. One of possible approach can be bisection
method. If error of section is lower than selected threshold, no further
division occurs. Non-control points are set as empty value and image
is stored. In a reconstruction, missing values are computed from a
cubic spline interpolation. The proposed technique is tested on RGB
images, but same approach also works in grayscale. 

Quadtree based structures can also be used for image compression.
One of the first algorithms with this approach was presented in \cite{Shu94}.
In \cite{Kei09} improved algorithm was used. Quadtree leaf is further
divided according to its analysis (high or low detail image block).
Analysis is done using quantized histogram of difference block (mean
is subtracted from each block value). Peaks bigger than threshold
are found in histogram. Number of peaks is used as block classificator.
If block is low-detailed, its mean value is stored, otherwise division
continue. If block is 4x4, no further division is performed. Block
is stored with binary mask instead. The best mask is chosen from predefined
patterns. 

Dictionary based methods are well known in a lossless image compression
(PNG, GIF). Its modification for lossy compressions was proposed in
\cite{Dud07}. Basic ideas is to create longer sequences of the same
value using quantization. This can be best shown on simple example.
Assuming data sequence 159, 160, 159, 157, 170, we quantize this and
obtain 159, 159, 159, 159, 170. This can be better compressed using
dictionary method. Second option is to used similarity of blocks.
If difference error of two blocks is lower than threshold, block can
be replaced one with another. That also improves compression. Dictionary
based methods are similar to near-lossless ones and can also be used
that way.

\section{Comparison}

In this section, comparison of earlier discussed methods is conducted.
Compressed size is expressed in bits per pixel (\emph{bpp}). Using
this metric is more readable, because it is image dimension independent.
This value can be very easily computed using equation \ref{eq:BPP}.

\begin{equation}
bpp=\frac{8}{M\cdot N}\cdot size\label{eq:BPP}
\end{equation}

$M$and $N$ refer to image dimensions, $size$ is size of compressed
image in bytes.

Besides compressed size, visual quality of reconstructed image is
very important for lossy compressions. As one of standards in this
area is PSNR (Peak signal-to-noise ratio). We can found results using
this metric in almost every article dealing with lossy image compression.
PSNR is computed based on MSE (Mean Square Error). Full equation is
shown in \ref{eq:PSNR}.

\[
MSE=\frac{1}{M\cdot N}\sum_{i=0}^{M}\sum_{j=0}^{N}||I(i,j)-K(i,j)||^{2}
\]

\begin{equation}
PSNR=10\cdot log\frac{255^{2}}{MSE}\label{eq:PSNR}
\end{equation}

$M$and $N$ refer to image dimensions, $I(i,j)$ signifies original
pixel intensity at location (i,j) and $K(i,j)$ denotes pixel intensity
in reconstructed image.

\subsection{Lossless compression}

\begin{figure}[H]
\begin{centering}
\includegraphics[scale=0.2]{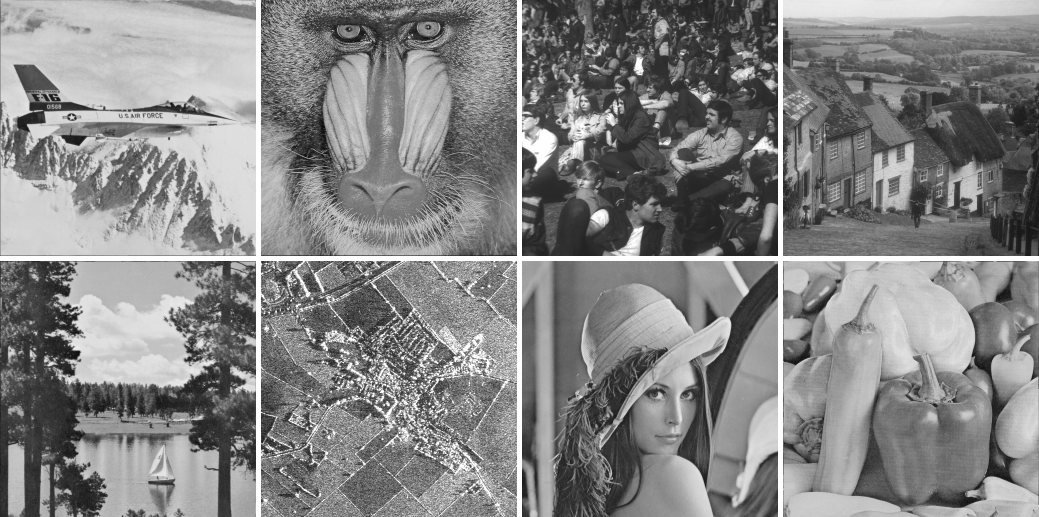}
\par\end{centering}

\caption{Test images used in \prettyref{tab:LOSSLESS_TEST}. From left to right:
Airplane, Crowd, Baboon (or Mandrill), Goldhill, Lake, Landsat, Lenna,
Peppers. \label{fig:LOSSLESS_IMAGES}}
\end{figure}

\begin{figure}[H]
\includegraphics[scale=0.4]{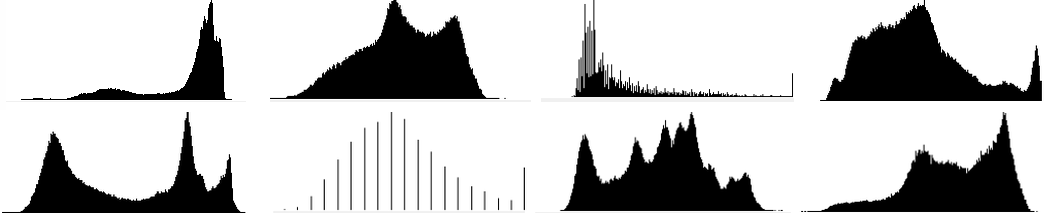}\caption{Histograms of test images from \prettyref{fig:LOSSLESS_IMAGES}. The
order of images is same as in \prettyref{fig:LOSSLESS_IMAGES}. \label{fig:LOSSLESS_HISTOGRAMS}}

\end{figure}

Lossless compression was tested on the image set from \prettyref{fig:LOSSLESS_IMAGES}
(their histograms can be found on \prettyref{fig:LOSSLESS_HISTOGRAMS}).
Results are taken directly from cited articles and compared to each
other. Same methods can be often found in different articles for comparison
with their methods. We collect all those data and created summarized
table. Sometimes, tests for certain image were not provided, in that
case value ``$-$'' is used. Collected results can be seen in \prettyref{tab:LOSSLESS_TEST}.

\begin{table}[H]
\centering{}%
\begin{tabular}{|c|c|c|c|c|c|c|c|c|c|c|c|c|c|}
\hline 
 & {\small \cite{Cro95}} & {\small \cite{Sai96}} & {\small \cite{Wu96}} & {\small \cite{Mot00}} & {\small \cite{Wei00}} & {\small \cite{Rab02}} & {\small \cite{Pan07}} & {\small \cite{Bra09}} & {\small \cite{Kik09}} & {\small \cite{Put11}} & {\small \cite{Avr10}} & {\small \cite{Ala12}} & {\small \cite{Li13}}\tabularnewline
\hline 
\hline 
{\small Airplane} & {\small 5.69} & {\small 3.97} & {\small 3.51} & {\small 3.62} & {\small 3.78} & {\small 3.98} & {\small 3.92} & {\small 4.03} & {\small 3.90} & - & {\small 4.23} & {\small 4.39} & -\tabularnewline
\hline 
{\small Baboon} & {\small 7.30} & {\small 6.16} & {\small 5.74} & {\small 5.65} & {\small 6.04} & {\small 6.11} & {\small 5.93} & {\small 5.98} & {\small 6.04} & {\small 5.08} & {\small 6.88} & - & {\small 3.60}\tabularnewline
\hline 
{\small Crowd} & {\small 5.84} & - & {\small 3.76} & {\small 4.03} & {\small 3.91} & {\small 4.19} & - & - & - & - & - & - & -\tabularnewline
\hline 
{\small Goldhill} & {\small 6.66} & {\small 4.63} & {\small 4.63} & {\small 4.66} & {\small 4.71} & {\small 4.84} & {\small 4.54} & - & {\small 4.48} & - & {\small 4.73} & {\small 5.08} & {\small 4.30}\tabularnewline
\hline 
{\small Lake} & {\small 6.84} & - & {\small 4.90} & {\small 4.93} & {\small 4.92} & {\small 5.90} & - & - & - & {\small 8.53} & {\small 5.37} & {\small 5.35} & -\tabularnewline
\hline 
{\small Landsat} & {\small 3.98} & - & {\small 3.99} & {\small 4.04} & {\small 7.17} & {\small 7.36} & - & - & - & - & - & - & -\tabularnewline
\hline 
{\small Lenna} & {\small 6.84} & {\small 4.18} & {\small 4.11} & {\small 4.05} & {\small 4.23} & {\small 4.31} & {\small 4.15} & {\small 4.44} & {\small 4.24} & {\small 5.07} & {\small 4.54} & {\small 5.09} & {\small 3.80}\tabularnewline
\hline 
{\small Peppers} & {\small 5.34} & {\small 4.68} & {\small 4.20} & {\small 4.16} & {\small 4.85} & {\small 4.93} & {\small 4.55} & {\small 4.59} & {\small 4.51} & - & {\small 4.81} & {\small 5.30} & {\small 2.80}\tabularnewline
\hline 
\textbf{AVG} & \textbf{6.06} & \textbf{4.91} & \textbf{4.50} & \textbf{4.54} & \textbf{4.95} & \textbf{5.20} & \textbf{4.80} & \textbf{4.76} & \textbf{4.63} & \textbf{4.42} & \textbf{5.21} & \textbf{5.04} & \textbf{3.63}\tabularnewline
\hline 
\end{tabular}\caption{Lossless compression sizes in bpp for standard set of images with
resolution \emph{512x512}. PNG is \cite{Cro95}, CALIC is \cite{Wu96},
JPEG-LS is \cite{Wei00} and JPEG2000 is \cite{Rab02}. Other methods
are not acknowledged as standards. \label{tab:LOSSLESS_TEST}}
\end{table}

Even though there are some missing data, we can see the overall compression
performance. The worst overall result came from the oldest tested
algorithm, PNG (\cite{Cro95}). The only image, where this method
is competitive, is landsat. Its because of its character - image has
only 20 brightness values (see \prettyref{fig:LOSSLESS_HISTOGRAMS}).
On the other hand, the best result is from the newest one, that is
using DWT transform (\cite{Li13}). 

There can also be seen some kind of dependency between compressed
size and histogram. If we look at image of Airplane and Crowd, they
have very similar histograms (only reversed). Their compression ratios
are also very similar for all tested methods.

\subsection{Lossy compression}

\begin{figure}[H]
\begin{centering}
\includegraphics[scale=0.4]{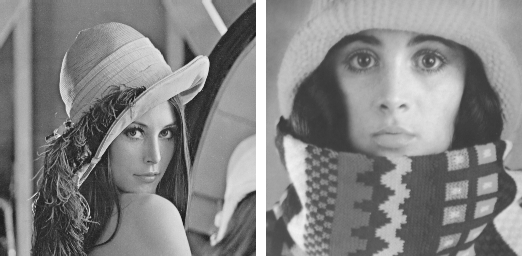}
\par\end{centering}

\caption{Test images used in \prettyref{fig:LOSSY_GRAPH_LENNA} and \prettyref{fig:LOSSY_GRAPH_TRUI}.
From left to right: Lenna, Trui \label{fig:LOSSY_IMAGES}}

\end{figure}

For the lossy compression test, standard image, that can be found
in practically every publication, is Lenna (mostly in resolution 256x256).
Our tests will be summarized in graph, because of its clarity. In
lossless test, there was only one possible result. For lossy compression,
we can get different PSNR values for different compressed size. In
our tests we used images from \prettyref{fig:LOSSY_IMAGES}.

\begin{figure}[H]
\begin{centering}
\includegraphics[bb=40bp 240bp 550bp 560bp,clip,scale=0.7]{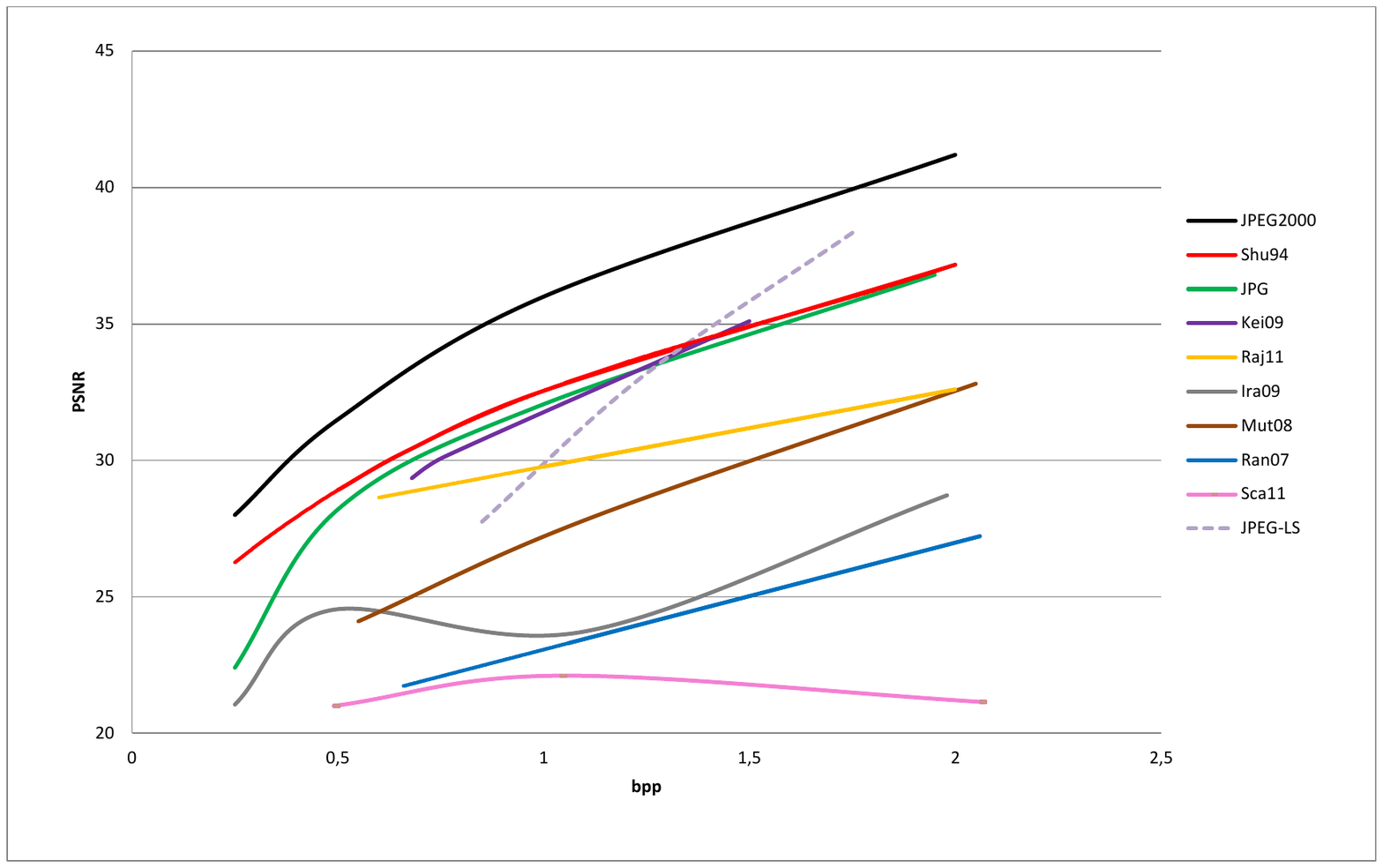}
\par\end{centering}

\caption{PSNR and compression in bpp for image Lenna in resolution of 256x256.
Graph legend is sorted from the best to worst overall method.\label{fig:LOSSY_GRAPH_LENNA}}

\end{figure}

First test is using Lenna image in resolution 256x256. Results can
be seen on \prettyref{fig:LOSSY_GRAPH_LENNA}. Best result were achieved
with JPEG2000 compression, that significantly outperforms other methods.
Very similar results were achieved with JPG and both tested quadtree
methods (\cite{Shu94} and \cite{Kei09}). For larger compression
ratios, neural network from \cite{Raj11} is also competitive. The
worst results were, as aspected, achieved with contour based compression
(\cite{Sca11}). This method is more suitable for specialized images,
mostly cartoon graphic (as was discussed in \cite{Mai11}).

We also added one of near-lossless methods for comparison. That is
JPEG-LS, based on \cite{Wei00}. We set different quality loss and
simulated classic lossy method. As we can see, for higher bitrates,
the quality improves very fast, which is basic of near-lossless method.

We can see some strange behaviour for neural network based compression
(\cite{Ira09}). For lower bitrate, we have better PSNR. This effect
can be caused by learning problems with neural networks. We have tried
to bypass this problem by taking the best results from several independent
tests, but as we can see, problem still persist in a lower bitrates.

\begin{figure}[H]
\begin{centering}
\includegraphics[bb=40bp 240bp 550bp 560bp,clip,scale=0.7]{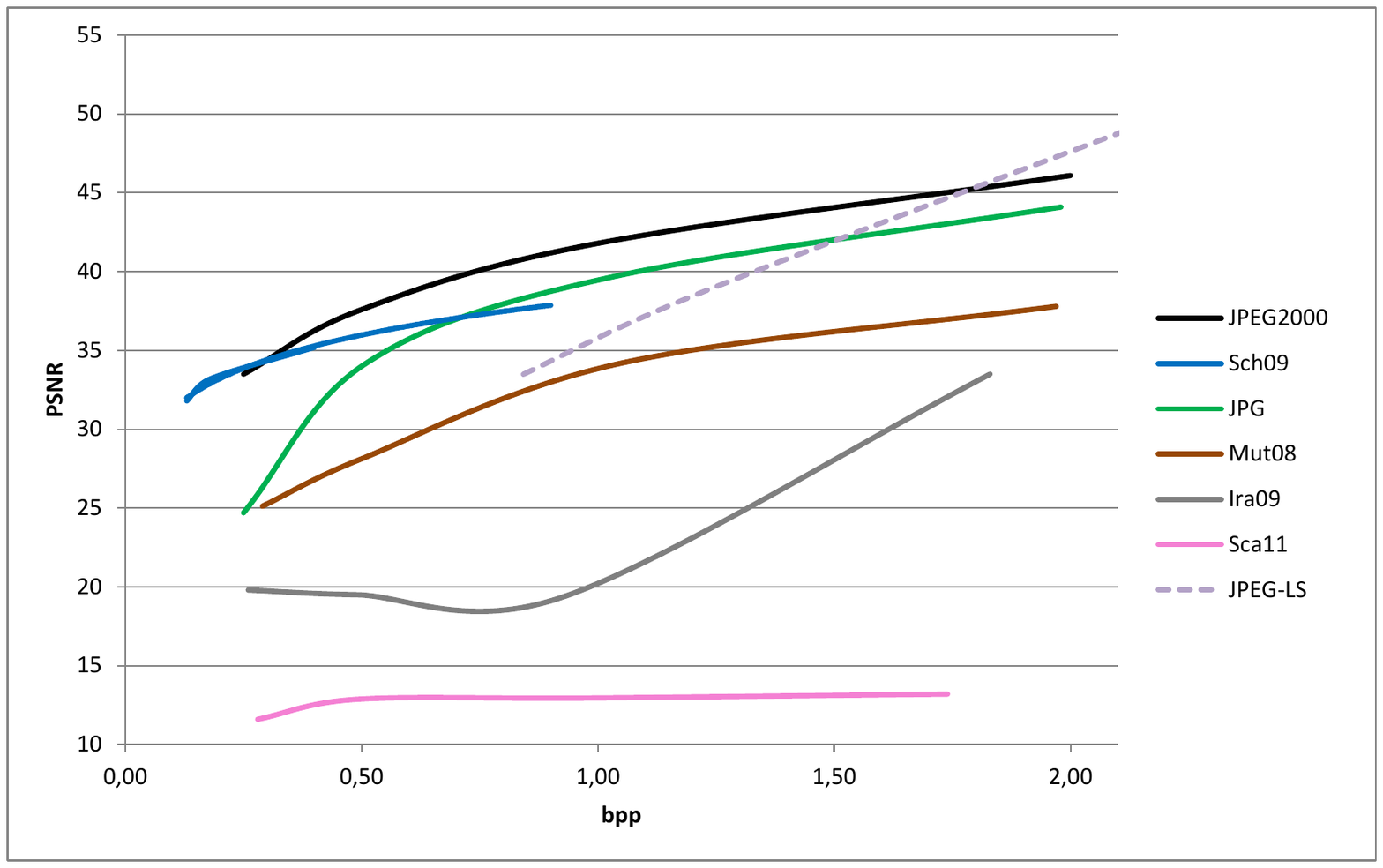}
\par\end{centering}

\caption{PSNR and compression in bpp for image Trui in resolution of 256x256.
Graph legend is sorted from the best to worst overall method.\label{fig:LOSSY_GRAPH_TRUI}}

\end{figure}

Second test is using Trui image in resolution 256x256. Results can
be seen on \prettyref{fig:LOSSY_GRAPH_TRUI}. Results for our second
image are similar to first one. The best method is again JPEG2000,
but this time with lower lead. Improvement for lower bitrates of JPEG2000,
\cite{Sch09}, provides better results as expected. However, for larger
bitrates, quality can be outperformed by JPG. Spline based method
\cite{Mut08} provides better results, than for Lenna image. If we
look at Trui, we can see the consequence of this behaviour. Trui has
more smooth areas, and spline can better describe them. Lenna has
more hard edges, were spline is not quite accurate. 

Test with near-lossless JPEG-LS (based on \cite{Wei00}) behave similar
to one with Lenna image. Only for now, the progress of quality loss
is slower. The reason is similar to one, why spline methods works
better.

The worst result is, again, achieved with \cite{Sca11}.

\section{Conclusion}

We have discussed and compared various image compression methods.
There are many more techniques, that were not covered. The need for
a better quality and compression ratio is very important. However,
as we can see today, compression techniques commonly used for images
are not very modern. The question of compression is much more complicated. 

New compression algorithm is just the beginning. Second step, to carry
through that method, is much more difficult. We can take simple example
from our comparison. Prediction based method (JPEG-LS) has been developed
in year 2000 and it is recognized as industrial standard. It outperforms
PNG and lossless JPEG2000 (in both, quality and speed). And how common
can we found images in this format?

\bibliographystyle{plain}
\bibliography{images_bibtex, neural_nets_bibtex}

\end{document}